\documentstyle[epsf]{article}
\huge
\renewcommand{\theequation}{\arabic{section}.\arabic{equation}} 
\def\setzero{\setcounter{equation}{0}}

\newcounter{eqalph}
\def\bph{\setcounter{eqalph}{\value{equation}}
   \addtocounter{eqalph}{1}
   \setcounter{equation}{0}
   \renewcommand{\theequation}{\arabic{section}.\arabic{eqalph}\alph{equation}}}
\def\eph{\setcounter{equation}{\value{eqalph}}
   \renewcommand{\theequation}{\arabic{section}.\arabic{equation}}
\par\noindent}

\begin{document}
\baselineskip 18pt
\newcommand{\ii}{\mbox{i}}
\def \sech{{\rm sech}}
\def \tanh{{\rm tanh}}
\def \cn{{\rm cn}}
\def \sn{{\rm sn}}
\def \max{{\rm max}}
\def \min{{\rm min}}
\def\bm#1{\mbox{\boldmath $#1$}}
\newfont{\bg}{cmr10 scaled\magstep4}
\newcommand{\bzr}{\smash{\hbox{\bg 0}}}
\newcommand{\bzl}{%
 smash{\lower1.7ex\hbox{\bg 0}}}
\title{KPZ  Equation 
and  
Surface Growth Model} 
\date{\today}
\author{ Masato {\sc Hisakado}  
\\
\bigskip
\\
{\small\it Department of Pure and Applied Sciences,}\\
{\small\it University of Tokyo,}\\
{\small\it 3-8-1 Komaba, Meguro-ku, Tokyo, 153, Japan}}
\maketitle

\vspace{20 mm}

Abstract

We consider the ultra-discrete Burgers equation. 
All variables of the equation are discrete.
We classify the equation  into five regions in the parameter space.
We discuss   behavior of solutions.
Using this equation 
we  construct the deterministic surface growth models respectively.
Furthermore we introduce  noise into the ultra-discrete 
Burgers equation.
We present the  automata models  of the KPZ equation.
One model corresponds to the discrete version of the 
ASEP and the other  to the Kim-Kosterlitz model.


\vfill
\par\noindent
{\bf  }

\newpage
\section{Introduction}

The noisy Burgers equation appears in a variety of problems 
in non-equilibrium statistical mechanics.\cite{b}
Denoting the velocity field by $u(x,t)$ the equation reads
\begin{equation}
\frac{\partial  u}{\partial t}=
\frac{\partial ^{2}u}{\partial x^{2}}
+2u\frac{\partial u}{\partial x}\;\;\;+{\rm  noise}.
\label{be}
\end{equation}
Burgers studied the noiseless equation with 
random initial data.
We are interested in the case  the noise is of the form 
 $\partial \xi/\partial x$, 
 to conserve $u$ locally.
 Then (\ref{be}) is a prototype for  a driven diffusive system.
If  we set   $h=\int u dx$, 
(\ref{be}) becomes 
\begin{equation}
\frac{\partial  h}{\partial t}=
\frac{\partial ^{2}h}{\partial x^{2}}
+(\frac{\partial h}{\partial x})^{2}+\xi.
\label{kpz}
\end{equation}
We can  interpret $h(x,t)$ as the height of a
(one dimensional) surface. 
(\ref{kpz}) is called the Karder-Parisi-Zhang (KPZ)   equation.\cite{kpz}
This equation  governs the shape  fluctuations of the 
various growth model.\cite{kk}

In this paper  we investigate the Burgers equation  and 
the surface growth model  without noise and with noise.
With noise, the Burgers equation describes steady 
growth in the long time limit, however, without noise, 
it becomes the relaxation  of an initially rough surface to the flat surface.


The relations between the Burgers equation and the 
surface growth model 
 as studied 
from the view point  of the universality classes, 
because the surface growth models are the discrete 
and the Burgers equation is continuous.
We study the direct relations between discrete and continuous 
using the ultra-discrete method.
Recently Nagatani, Nishinari and Takahahi presented 
the
ultra-discrete version of the Burgers equation .\cite{n},\cite{nt}
We study   solutions  of the this equation.
Behavior  of solutions can be classified into  
five  regions  in the parameter space.
The solutions can be understood from the viewpoint of 
creation and annihilation of 
particles and anti-particles.
Using  behavior  of  particles and anti-particles 
we can construct the surface models.
The surface model corresponds to the 
deterministic version of the 
KPZ equation.
Further more we present ultra-discrete 
version of the noisy 
Burgers equation and the KPZ equation.
We show the direct relations between 
the cellular automata  and KPZ equation.
One model of these cellular automata 
is   the time discrete  version of the 
asymmetric simple exclusion process (ASEP) 
which  belongs to KPZ universality class.
The other model corresponds to the restricted solid on solid (RSOS) model 
which is introduced by Kim and Kosterlitz.
\cite{kk} 

This paper is organized as follows.
In the section 2 we classify  the equation into five regions  
in the parameter space and discuss the behavior 
of the solutions from the viewpoint of 
particles and anti-particles.
In the section 3 we study the 
symmetry of the equation.
We explain why the automata model represent the property of 
continuous Burgers equation.
In the section 4 we discuss the 
stability of the equation.
We study the meaning of the ultra-discrete limit.
In the section 5 we construct the deterministic surface growth models.
In the section 6 we introduce the noise into the 
discrete Burgers equation.
In the section 7 we construct the noisy surface growth models using 
the ultra-discrete KPZ equation.
The models are equivalent to the well known models which 
belong to the KPZ universality.
The last section is devoted to the concluding remarks.

First we review the derivation of the ultra-discrete  Burgers equation.
Discretizing of both time and space variables 
of diffusion equation, we can obtain 
\begin{equation}
\frac{f_{j}^{t+1}-f_{j}^{t}}{\Delta T}=
\frac{f_{j+1}^{t}-2f_{j}^{t}+f_{j-1}^{t}}{(\Delta X)^{2}},
\label{dde}
\end{equation}
where $\Delta T$ and $\Delta X$  are lattice intervals 
in $x$ and $t$ respectively.
Using the  discrete  analogue of  Cole Hopf transformations 
\begin{equation}
u_{j}^{t}=c\frac{f_{j+1}^{t}}{f_{j}^{t}},
\end{equation}
where $c$ is constant. 
We can obtain the lattice  version of the 
Burgers equation
\begin{equation}
u_{j}^{t+1}=
u_{j-1}^{t}
\frac{1+\frac{1-2\delta}{c\delta}u_{j}^{t}+\frac{1}{c^{2}}
u_{j}^{t}u_{j+1}^{t}}
{1+\frac{1-2\delta}{c\delta}u_{j-1}^{t}+\frac{1}{c^{2}}
u_{j-1}^{t}u_{j}^{t}},
\label{dbe}
\end{equation}
where $\delta=\Delta T/(\Delta X)^{2}$.

We introduce a  parameter $\epsilon$  and new variables 
\begin{equation}
u_{j}^{t}=e^{U_{j}^{t}/\epsilon},\;\;\;
\frac{1-2\delta}{c\delta}=e^{-M/\epsilon},\;\;\;
c^{2}=e^{L/\epsilon}.
\label{udl}
\end{equation}
Then (\ref{dbe}) becomes 
\begin{eqnarray}
U_{j}^{t+1}=U_{j-1}^{t}&+&\epsilon \log[1+\exp(\frac{U_{j}^{t}-M}{\epsilon})
+\exp(\frac{U_{j}^{t}+U_{j+1}^{t}-L}{\epsilon})]
\nonumber \\
&-&\epsilon \log[1+\exp(\frac{U_{j-1}^{t}-M}{\epsilon})
+\exp(\frac{U_{j-1}^{t}+U_{j}^{t}-L}{\epsilon})]
\end{eqnarray}
Here we take the so called ``ultra-discrete limit''\cite{tt}
\begin{equation}
\epsilon \longrightarrow 0+.
\label{udl}
\end{equation}
Then we can obtain 
\begin{equation}
U_{j}^{t+1}=U_{j}^{t}+\min (M,U_{j-1}^{t},L-U_{j}^{t})-
\min (M,U_{j}^{t},L-U_{j+1}).
\label{udbe}
\end{equation}
Here we use the relation 
\begin{equation}
\lim_{\epsilon \rightarrow 0+}
\epsilon 
\log(e^{\frac{A}{\epsilon}}+e^{\frac{B}{\epsilon}})
=
\max(A,B).
\label{max}
\end{equation}
We call (\ref{udbe}) the ultra-discrete  Burgers equation.

\setzero
\section{Classification}
\begin{figure}[htb]
  \begin{center}
    \leavevmode
           \epsfxsize = 8cm 
       \epsfbox{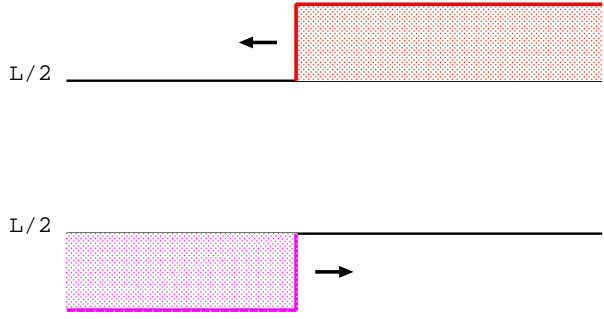}
     \end{center}  
\caption{ Shock wave solution }
\label{f1}
\end{figure}

(\ref{udbe}) has a  solution 
\begin{equation}
U_{j}^{t}=\frac{L}{2}+\max(0,K(j+1)+\Omega t+\Theta_{0})
-\max(0,K(j)+\Omega t+\Theta_{0}),
\label{solution}
\end{equation}
with
\begin{eqnarray}
\Omega=
\left\{\begin{array}{ll}
|K|&\mbox{for $M\geq\frac{L}{2},$}\\        
\max(\frac{L}{2}-M,K,-K)-\frac{L}{2}+M
& \mbox{for $M<\frac{L}{2},$}
\end{array}
\right.
\end{eqnarray}
where $\Theta_{0}$ is constant.
This solution  corresponds to the famous  shock wave solution 
of the Burgers equation.Fig.\ref{f1}

Here we classify five cases.
\begin{eqnarray}
&(A1)&\;\;\frac{L}{2}+|K|\leq M,\;\;\;(A2)\;\;
\frac{L}{2}<M<\frac{L}{2}+|K|,
\nonumber \\
&(B)&\;\;\frac{L}{2}=M,\;\;\;
\nonumber \\
&(C1)&\;\;\frac{L}{2}>M>\frac{L}{2}-|K|,\;\;\;
(C2)\;\;\frac{L}{2}-|K|\geq M.
\end{eqnarray}

At first we study the case (A) and (B).
Note that in this case the dispersion relation is $\Omega=|K|$.
The sign of $K$ chooses  the propagating
 direction and the velocity of the  wave  is  $-1$ $(+1)$ for $K>0$  $(K<0)$.
If we set  $U_{j}^{t}\geq\frac{L}{2}$, then  (\ref{udbe}) becomes
$U_{j}^{t+1}=U_{j+1}^{t}$.
On th other hand,  setting $U_{j}^{t}\leq\frac{L}{2}$, (\ref{udbe}) becomes
$U_{j}^{t+1}=U_{j-1}^{t}$.  
In both cases equations become  linear.
Then stable rectangular wave becomes a solution.
We can write the rectangular  solution explicitly as
\begin{eqnarray}
U_{j}^{t}
&=&\frac{L}{2}+\max(0,K(j+1)+\Omega t+\Theta_{0})
-\max(0,K(j)+\Omega t+\Theta_{0})
\nonumber \\
& &
-\max(0,K(j+1+l)+\Omega t+\Theta_{0})
+\max(0,K(j+l)+\Omega t+\Theta_{0}).
\label{solution2}
\end{eqnarray}
The size  of the rectangular is   $K \times l$ Fig.2.

 \begin{figure}[htb]
  \begin{center}
    \leavevmode
           \epsfxsize = 8cm 
       \epsfbox{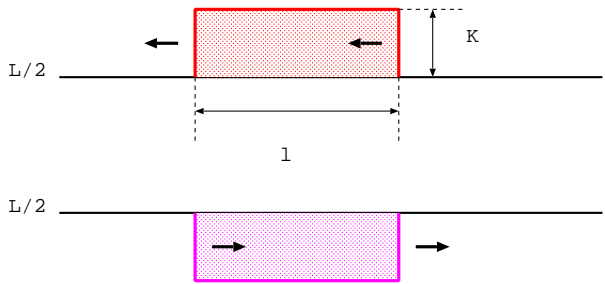}
     \end{center}  
\caption{ Rectangular solution}
\label{2g}
\end{figure}

Here we introduce an  interpretation of  particles and anti-particles.
We call that  particles  (anti-particles) are  created at time $t$ and $j$ 
when $U_{j}^{t}>\frac{L}{2}$ ($U_{j}^{t}<\frac{L}{2}$).
$U_{j}^{t}-\frac{L}{2}$ can be understood as the number  of 
particles or an anti-particles.
In this mean  this particles  or anti-particles  are   Boson.
Then  we call $U_{j}^{t}=\frac{L}{2}$ 
as vacuum  state.
From this interpretation 
this equation has chairarity.
Particles go only to the left  and anti-particles go only to the right.
If there are only 
particles (anti-particles), 
creation and annhiration  of particles (anti-particles) 
do not occur.
It  can be seen from the result (\ref{solution2}) is  
stable solution of (\ref{udbe}).

\begin{figure}[htb]
  \begin{center}
    \leavevmode
           \epsfxsize = 8cm 
       \epsfbox{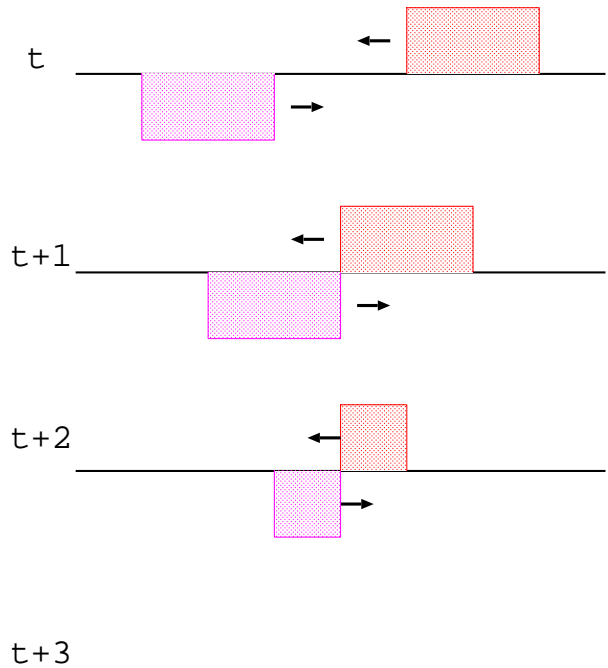}
     \end{center}  
\caption{Collision of particles and anti-particles}
\label{3g}
\end{figure}

We need to study interaction of particles and anti-particles.
To consider interaction we 
consider collision of particles and anti-particles.
Annihilation of particles and 
 anti-particles  occurs.
If same numbers of particles and 
anti-particles  meet, 
they become vacuum.
The configuration of particles and anti-particles 
 are in  Fig.\ref{3g}.
If the number  of particles (anti-particles) are larger than the number of 
particles of anti-partials (particles),
then only particles (anti-particles) whose number is difference of 
particles and anti-particles   survive.

Note that the point where particles and anti-particles meet 
does not move.
It can be seen also in continuous limit as the collision of the shock waves.
In the Burgers equation (\ref{be})
setting $u(x,t)=u(x-vt)$, where $v$ is propagation velocity.
Imposing the boundary condition 
$u\rightarrow u_{\pm}$ and $\partial u\rightarrow 0$ for $x\rightarrow \pm 
\infty$, then we can obtain the soliton condition 
\begin{equation}
u_{+}+u_{-}=-v
\label{sc}
\end{equation}
In the limit $u_{+}=u_{-}$ this condition 
implies $v=0$.  It means that the collision point does not move  in 
the continuous Burgers equation.
In fact the number  of particles (anti-particles) is larger than 
that  of anti-particles (particles),  
the collision point  moves 
 to the left (right) and the particles 
(anti-particles) only survive.
It can be seen in the soliton condition (\ref{sc}).

The other configuration is in Fig.4.
In the case (A) 
creation of  pairs of particles and anti-particles between 
particles and anti-particles occurs.
On the  other hand 
in the case (B) particles and anti-particles 
part over without  creation of pairs of particles and anti-particles.
It is the crucial difference between (A) and (B).

We consider the case same number $K$ of particles and anti-particles 
part over. 
In the case (A1)  $K$ particles and $K$ anti-particles are 
 created.
On the other hand  
in the case (A2)  creation of $M-\frac{L}{2}$  pairs of  particles and anti-particles 
  occurs.
If the number of particles are $K_{1}$ and anti-particles  are 
$K_{2}$, 
in the case (A1)
$K$ pairs  are created, 
in the case  (A2)creation of 
$M-\frac{L}{2}$  pairs occurs, where
min$(K_{1},K_{2})=K$.

\newpage
\begin{figure}[htb]
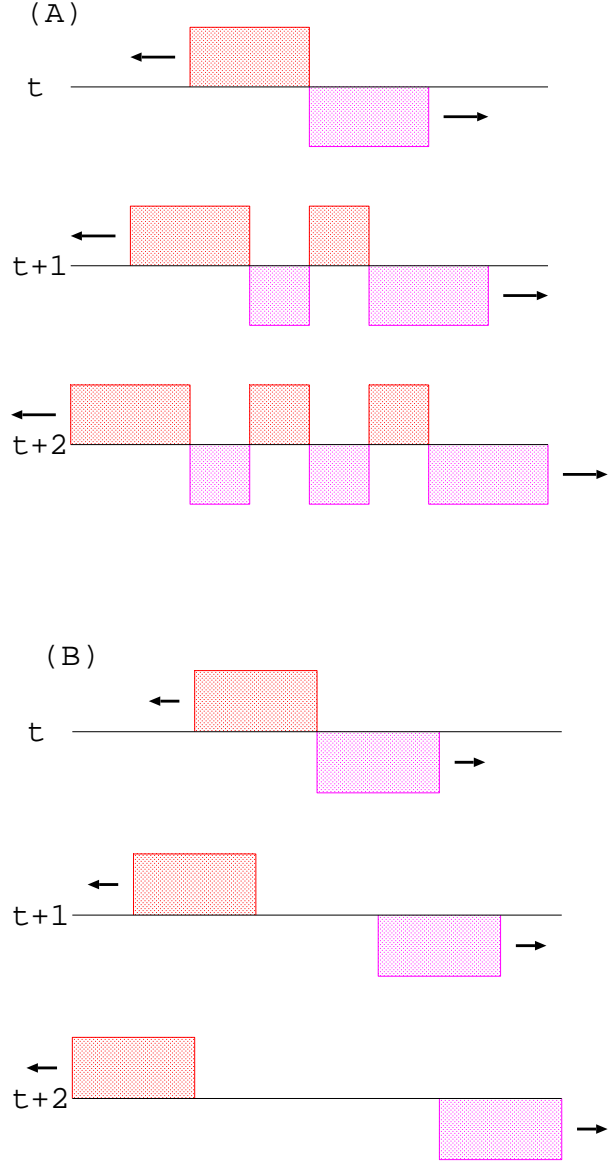

  \begin{center}
    \leavevmode
           \epsfxsize = 8cm 
       \epsfbox{kpz4a.eps}
\end{center}  

\vspace{1cm}
\begin{center}
    \leavevmode
           \epsfxsize = 8cm 
       \epsfbox{kpz4b.eps}
\label{4g}
\end{center}

\caption{Separations  of particles and anti-particles 
in the cases  (A) and  (B)}
\end{figure}

Next we consider the case (C).
We can rewrite  the dispersion relations 
\begin{equation}
(C1)
\;\;\;
\Omega=K+M-\frac{L}{2},\;\;\;
(C2)
\;\;\;
\Omega=0.
\label{dr}
\end{equation}
The characteristic  is the dispersion relations 
with  a gap in the case (C1).
From (\ref{dr}) 
in the case (C2)
 the shock wave solution (\ref{solution}) 
does not move.
In other words $U_{j}^{(t+1)}=U_{j}^{(t)}$.
Then any initial configurations of particles and anti-particles 
does  not move. 
In the case (C1) the speed of the 
shock wave solution is
\begin{equation}
v=\frac{|K|+M-\frac{L}{2}}{K},
\end{equation}
where 
$v$ is the speed of the shock wave.
The shock wave with positive parity, $K>0$, 
propagates left with negative velocity, whereas the shock wave with 
opposite parity, $K<0$, propagate 
in the forward direction 
as we have shown  in the cases  (A) and (B).
However in the case (C1)
the speed $v$ has a range.
The maximum  speed  is $1$ or $-1$.
Notice that in the limit $K\rightarrow \pm \infty$ 
the speed of the shock wave become  $\pm 1$.
If the shock waves 
with  the positive parity and negative parity
collide, 
the annihilation of the particles and anti-particles occurs
as we have seen in the cases  (A) and (B).
The difference is the speed of the annihilation. 
In the case (C1)
there are  not rectangular solutions. Then the 
pair creation of particles and anti-particles 
does not occur.
{}From these results 
the case (C1)  represents  the property of the 
continuous Burgers equation.

\setzero
\section{Symmetry of Equations}

Here we note that the Burgers equation (\ref{be}) 
is invariant under the parity transformation
$x\rightarrow -x$ provided  $u\rightarrow -u$.
This feature is related to the presence 
on a single spatial derivative in the derivative term.
We can not see this parity invariant 
in the discrete version of the 
Burgers equation.
We consider the other ultra-discrete limit
\begin{equation}
\epsilon \longrightarrow -0.
\label{ud1}
\end{equation}
In this limit we use  a following relation
\begin{equation}
\lim_{\epsilon \rightarrow 0-}
\epsilon 
\log(e^{\frac{A}{\epsilon}}+e^{\frac{B}{\epsilon}})
=
\min(A,B).
\label{min}
\end{equation}
Using this relation we 
can take the other ultra-discrete limit. Then  we can obtain 
\begin{eqnarray}
U_{j}^{t+1}
&=&U_{j}^{t}+\max (M,U_{j-1}^{t},L-U_{j}^{t})-
\max (M,U_{j}^{t},L-U_{j+1}^{t})
\nonumber \\
&=&
U_{j}^{t}+\min (\tilde{M},U_{j+1}^{t},\tilde{L}-U_{j}^{t})-
\min (\tilde{M},U_{j}^{t},L-U_{j-1}^{t}),
\label{udbe1}
\end{eqnarray}
where 
$\tilde{L}=L$ and $\tilde{M}=L-M$.
(\ref{udbe1}) can be obtained  using the transformations of (\ref{udbe})
\begin{equation}
j+1\longrightarrow j-1,\;\;\;
j-1\longrightarrow j+1,\;\;\;
M   \longrightarrow \tilde{M},\;\;\;
L   \longrightarrow \tilde{L}.
\label{xsy}
\end{equation}
This is the parity transformation.
(\ref{xsy}) corresponds to $x\rightarrow -x$ in the continuous case.
(\ref{ud1}) is the transformation 
$U_{j}^{t}\longrightarrow -U_{j}^{t}$.
Then the invariance under parity transformation 
 is recovered.
The particles move  only to the right and 
the anti-particles only to the left 
for (\ref{udbe1}).
The particles and anti-particles move to the 
opposite  directions  to (\ref{udbe}).

Here we consider why the parity invariance  can be seen 
in the continuous case (\ref{be}) and ultra-discrete limit (\ref{udbe}), 
but not be seen in the full discrete type euqtaion (\ref{dbe}).
The relation between the discrete variables and the 
continuous variables are following
\begin{equation}
u(j\Delta x, t\Delta T)=\frac{1}{\Delta X}
\log \frac{u_{j}^{t}}{c}.
\label{im}
\end{equation}
Using (\ref{udl}),  we can obtain a relation 
\begin{equation}
u(j\Delta x, t\Delta T)=\frac{1}{\Delta X \epsilon}
(U_{j}^{t}-\frac{L}{2}).
\label{l1}
\end{equation}
The vacuum state $U_{j}^{t}=L/2$ corresponds 
to $u=0$.
It can be seen  in the collision  of the shocks.

Using the relation (\ref{im}) and (\ref{l1})
the relation among continuous,  discrete  and 
ultra discrete variables are 
\begin{equation}
v\sim \log u_{j}^{t}\sim
U_{j}^{t}.
\label{kankei}
\end{equation}
Then the the parity invariance 
is the same
in the continuous (\ref{be})  and 
ultra-discrete Burgers equations (\ref{udbe}).

\setzero
\section{Stability}
Here we consider  stability of the 
solution  of the discrete Burgers equation (\ref{dbe}).
We substitute  plane wave solution $u_{j}^{t}=e^{{\rm i}jk+\omega t}$
 into the discrete diffusion equation (\ref{dde}).
We can obtain the condition for the stability condition $|e^{\omega}|\le 1$,
\begin{equation}
\delta=0.
\label{sta}
\end{equation}
Using the second and the third equations of (\ref{udl})
we can get the relation
\begin{equation}
\delta=\frac{1}{2+\exp[\frac{1}{\epsilon}(\frac{L}{2}-M)]}
\end{equation}
Using this relation, 
the cases (A), (B) and (C)  correspond to $\delta=1/2$,
$\delta=1/3$ and $\delta=0$ respectively.Fig.5.
From the condition (\ref{sta}) 
the cases (A) and (B) 
are unstable and (C) is only stable in the
discrete Burgers equation.
It is the reason 
why 
the case (C) represents the property of the 
continuous Burgers equation (See section 2).
It also can be seen 
that we need to  set $\delta\rightarrow 0$ 
to obtain the continuous Burgers equation (\ref{be})  
from the lattice Burgers equation (\ref{dbe}).
On the other hand
 there is the rectangular solutions in the cases  (A) and (B)
which can not be seen in the 
Burgers equation.
The rectangular solutions of  (A) and (B)
correspond to  solutions of configurations of
two  solitons with opposite parity in the continuous Burgers equation.
To construct this configuration 
we need the condition 
that the solitons are well separated and 
non overlapping.
Then the size of the solutions are $\infty \times \infty$.
They are rescaled and becomes the rectangular solutions 
of the cases (A) and (B) in the ultra discrete limit.

\begin{figure}[htb]
  \begin{center}
    \leavevmode
           \epsfxsize = 8cm 
       \epsfbox{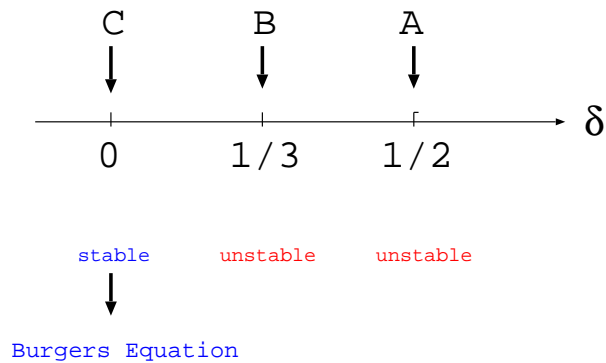}
     \end{center}  
\caption{Stability of discrete Burgers equation}
\end{figure}

\setzero
\section{Deterministic ultra-discrete KPZ equation}
If we set $u(x,t)=(\partial /\partial x)h(x,t)$, 
$h(x,t)$  is governed by the deterministic 
KPZ  equation.
To construct  automata model for  deterministic  KPZ equation,
from (\ref{kankei}) we can set
\begin{equation}
U_{j}^{t}=H_{j+1}^{t}-H_{j}^{t}+\frac{L}{2}.
\label{defh}
\end{equation}
The last term is selected the constant for the vacuum  state.
We rewrite (\ref{udbe}) 
\begin{equation}
H_{j}^{t+1}=\max(H_{j-1}^{t},H_{j}^{t}+\frac{L}{2}-M,
H_{j+1}^{t})+C,
\label{kpz1}
\end{equation}
where $C$ is a constant.
In the same way we can rewrite (\ref{udbe1}) 
by the independent variables $H_{j}^{t}$,
\begin{equation}
H_{j}^{t+1}=\min(H_{j-1}^{t},H_{j}^{t}+\frac{\tilde{L}}{2}-\tilde{M},
H_{j+1}^{t})+\tilde{C},
\label{kpz2}
\end{equation}
where 
$\tilde{C}$ is  a constant.
We call these  equations (\ref{kpz1}) and (\ref{kpz2}) 
``ultra-discrete deterministic KPZ equation.''

To describe  interface dynamics  we consider the
periodic condition
\begin{equation}
H_{1}^{t}=H_{N}^{t}.
\label{pc}
\end{equation}
Here $H_{i}^{t}$ denotes the height of the surface at the lattice 
site $i$ at the integer time $t$.
Solutions of  ultra-discrete KPZ equation 
are classed by (A1)-(C2) as the ultra-discrete Burgers equation.

\begin{figure}[htb]
  \begin{center}
    \leavevmode
           \epsfxsize = 8cm 
       \epsfbox{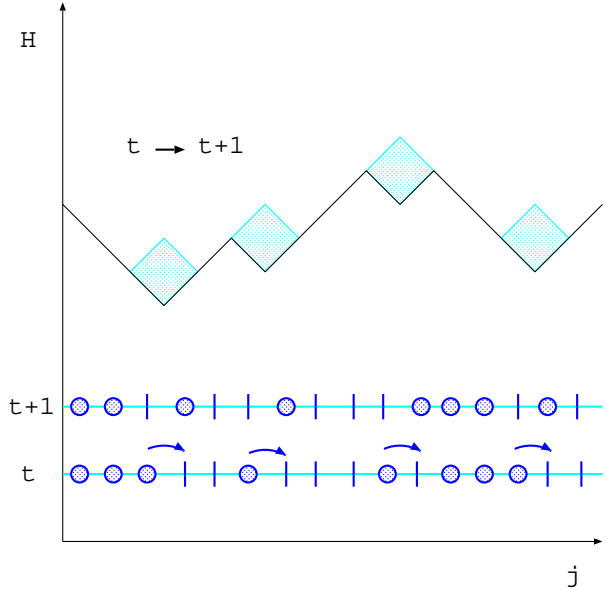}
     \end{center}  
\caption{Deterministic Surface Growth model A}
\label{6g}
\end{figure}

We consider  the  deterministic dynamical model.
In this interface model the integer height variable 
$H_{j}^{t}$ may differ on  neighbor sites $j$ and $j+1$ 
only $\pm 1$.
The two elementary  steps,  deposition and evaporation, 
which define the surface evolution are illustrated in 
Fig.\ref{6g}.
In a lattice gas language the height differences between neighorbouring 
sites are mapped to a 
particle occupation number $n_{x}^{t}=0,1$ with the presence 
of a particle on $x$ corresponding 
to slope $-1$ between sites $j-1$ and $j$ in the interface 
model and a vacancy  at  site $x$ corresponding to 
slope $+1$.
The particles have a simple dynamics.
With rate 1 they jump to the right (left)  except 
when the final site is occupied, in which case they stay
 (hard core exclusion).
In  this model growth (evaporate)  can occur   in local minima (maxima).
These models are deterministic.
The configuration at time $t+1$ is determined from  
the configuration  at time $t$ by a local rule which 
depends on nearest neighbors only.
The corresponding lattice gas evolves according to the automaton 
rule 184.\cite{w}


In the case (A1) we set  $K=\pm 1$, $C=-1$, $\tilde{C}=1$, 
in (\ref{kpz1}) and (\ref{kpz2}) 
describe following  deterministic model,
\bph 
\begin{equation}
H_{j}^{t+1}=\max(H_{j-1}^{t},H_{j+1}^{t})-1,
\label{1}
\end{equation}
and 
\begin{equation}
H_{j}^{t+1}=\min(H_{j-1}^{t},H_{j+1}^{t})+1.
\label{2}
\end{equation}
\eph
The model particles move only to the right 
corresponds to (\ref{1}) and 
the model only to the left 
is (\ref{2}).
The above deterministic  dynamical model is equivalent to the
ultra-discrete KPZ equation (\ref{1}) or  (\ref{2}).
(\ref{1}) is the surface  evaporate model and
(\ref{2}) is the surface growth model.
We call these models ``A model''.

We consider another interface model.
In this interface model the integer height variable 
$H_{j}^{t}$ may differ on  neighbor sites $j$ and $j+1$ 
only $\pm 1$or $0$.
Consider a one dimensional surface configuration parallel to 
the $j$ axis of a square lattice.

In the case (B) we set  $K=\pm 1,0$, $C=0$, in
(\ref{kpz1}) 
describe following  deterministic model,
\begin{equation}
H_{j}^{t+1}=\max(H_{j-1}^{t},H_{j}^{t},H_{j+1}^{t}).
\label{3}
\end{equation}
The deterministic model (3) was studied in \cite{ks}.

In the case (B) we set  $K=\pm 1,0$, $C=-1$ and $\tilde{C}=1$, in
 (\ref{kpz1}) and  (\ref{kpz2})
\bph
\begin{equation}
H_{j}^{t+1}=\max(H_{j-1}^{t},H_{j}^{t},H_{j+1}^{t})-1.
\label{4}
\end{equation}
and 
\begin{equation}
H_{j}^{t+1}=\min(H_{j-1}^{t},H_{j}^{t},H_{j+1}^{t})+1.
\label{5}
\end{equation}
\eph
In the model (\ref{4})
a particle is evaporated at the local maxima and 
in the model  (\ref{5})
a particle is deposited at the local minima.
The dynamical equations (\ref{4}) and (\ref{5})
restrict  the height difference $\pm 1$ or $0$ automatically.
We call the models  (\ref{4}) and (\ref{5})  `` B model ''.

\begin{figure}[htb]
  \begin{center}
    \leavevmode
           \epsfxsize = 8cm 
       \epsfbox{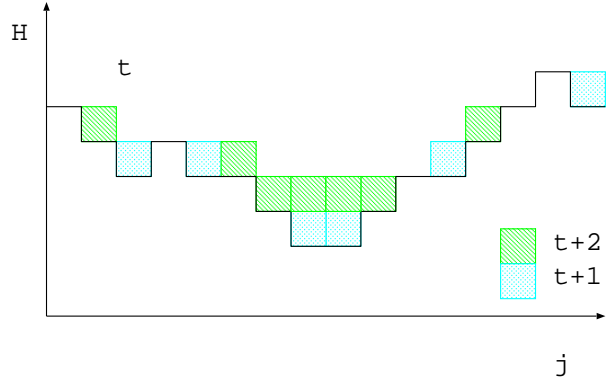}
     \end{center}  
\caption{Deterministic Surface Growth model B}
\end{figure}

In the case (C2) we also consider as the RSOS model.
If we 
 we set  $K=\pm 1,0$, $C=-1$ and $\tilde{C}=1$, in
 (\ref{kpz1}) and  (\ref{kpz2}), 
we can obtain the model 
whose dynamics is trivial 
$H_{j}^{t+1}=H_{j}^{t}+1$ or $H_{j}^{t+1}=H_{j}^{t}-1$.

\setzero
\section{Ultra-discretization with noise}
Next we consider the discrete Burgers equation with noise, 
\begin{equation}
u_{j}^{t+1}=
u_{j}^{t}
\frac{\frac{1}{u_{j}^{t}}+\frac{1-2\delta}{c\delta}+\frac{1}{c^{2}}
u_{j+1}^{t}+\frac{1}{c^{2}}\xi_{j+1}^{t}}
{\frac{1}{u_{j-1}^{t}}+\frac{1-2\delta}{c\delta}+\frac{1}{c^{2}}
u_{j}^{t}+\frac{1}{c^{2}}\xi_{j}^{t}},
\label{dbe1}
\end{equation}
where $\xi_{j}^{t}$ is the noise.
If we set $\xi_{j}^{t}=0$,  (\ref{dbe1}) is  the discrete Burgers equation 
(\ref{dbe}).
In the limit 
$\Delta X\rightarrow 0$ and $\Delta T\rightarrow 0$ 
with $\delta\rightarrow 0$,
(\ref{dbe1}) becomes the noisy Burgers equation
 \begin{equation}
\frac{\partial  u}{\partial t}=
\frac{\partial ^{2}u}{\partial x^{2}}
+2u\frac{\partial u}{\partial x}+\frac{\partial \xi(x,t)}{\partial x},
\label{ben}
\end{equation}
where the white noise 
$\xi (x,t)$
\begin{equation}
<\xi(x,t)\xi(x',t')>={\rm  D}\delta(x,x')\delta(t,t').
\end{equation}
The relation between the discrete noise and 
the continuous noise is 
\begin{equation}
\xi(x,t)=\frac{1}{(\Delta X)^{2}}
\log \xi_{j}^{t}.
\end{equation}
We introduce a parameter 
$\epsilon $ and new variables 
\begin{equation}
\xi_{j}^{t}=e^{X_{j}^{t}/{\epsilon}}.
\end{equation}
We take the ultra-discrete limit (\ref{udl})
in the discrete noisy Burgers equation (\ref{dbe1}).
We can obtain 
\begin{equation}
U_{j}^{t+1}=U_{j}^{t}+\max  (-M,-U_{j}^{t},U_{j}^{t}-L, X_{j+1}^{t}-
\frac{L}{2})-
\max (-M,-U_{j-1}^{t},U_{j}-L, X_{j}-\frac{L}{2}).
\label{udnbe}
\end{equation}

\setzero
\section{Ultra-discrete KPZ equation}
We rewrite this equation using the independent variables $H_{j}^{t}$ 
(\ref{defh})
\begin{equation}
H_{j}^{t+1}
=
\max (H_{j-1}^{t},H_{j}^{t}+\frac{L}{2}-M,H_{j+1}^{t},H_{j}^{t}+X_{j}^{t})
+C,
\label{nkpz1}
\end{equation}
where $C$ is is a constant.
This is the noisy version of the model (\ref{kpz1}).
In the same way we can obtain 
\begin{equation}
H_{j}^{t+1}
=
\min (H_{j-1}^{t},H_{j}^{t}+\frac{\tilde{L}}{2}-\tilde{M},
H_{j+1}^{t},H_{j}^{t}+\tilde{X}_{j}^{t})
+\tilde{C},
\label{nlpz2}
\end{equation}
where $\tilde{C}$ is is a constant.
This is the noisy version of the model (\ref{kpz2}).

Here we set $K=\pm 1$, $\tilde{C}_{1}=-1$, 
and $\tilde{C}_{2}=1$, we can obtain 
\bph
\begin{equation}
H_{j}^{t+1}
=
\max (H_{j-1}^{t},H_{j+1}^{t},H_{j}^{t}+X_{j}^{t})
-1,
\label{nkpz3}
\end{equation}
and 
\begin{equation}
H_{j}^{t+1}
=
\min (H_{j-1}^{t}, H_{j+1}^{t},H_{j}^{t}
+\tilde{X}_{j}^{t})
+1.
\label{nkpz4}
\end{equation}
\eph
These are the noisy version of the A models  (\ref{1}) and (\ref{2}).
Here we set the noise as following
\begin{eqnarray}
X_{j}^{t}=\tilde{X}_{j}^{t}=
\left\{
\begin{array}{ll}
1 & \mbox{probability  $q$ }\\
-1 & \mbox{probability  $1-q$}
\end{array}
\right.
\end{eqnarray}


In (\ref{nkpz3}) and (\ref{nkpz4}) we consider the 
non-deterministic dynamical model.
In the lattice gas language 
the particles have a simple dynamics.
With rate $q$   they jump to the right 
except when the final site is occupied 
for (\ref{nkpz3}).
For (\ref{nkpz4}) 
with rate $q$ they jump to the left.
In (\ref{nkpz3})  for $t$ time steps 
the probability distribution that  
a particle moves $n$ space steps 
is the binomial distribution
\begin{eqnarray}
P(t,n)=
\left(
\begin{array}{c}
t\\
n
\end{array}
\right )
q^{n}(1-q)^{t-n}.
\end{eqnarray}
It is well known that in the large $t$ limit 
with $\lambda=qt=$const, 
we can obtain the Poisson distribution.
Then in the  continuous limit of time 
we can obtain the 
Poisson process.

A lattice model of particles moving stochastically 
with hard-core exclusion is called  
the asymmetric  simple exclusion process (ASEP).\cite{l}
Each particle hops to the right (left) nearest site  
with the probability $p_{1}{\rm d}t$ ($p_{2}{\rm d}t$)
in every infinitesimal time interval ${\rm d}t$.
It is a Poisson process.
(\ref{nkpz3}) corresponds to the time discrete version of ASEP with 
$p_{1}=0$
 and (\ref{nkpz4}) corresponds  to the ASEP with $p_{2}=0$.



For the B model
we set $K=\pm 1,0$, $\tilde{C}_{1}=-1$, 
and $\tilde{C}_{2}=1$, we can obtain 
\bph
\begin{equation}
H_{j}^{t+1}
=
\max (H_{j-1}^{t},H_{j}^{t},H_{j+1}^{t},H_{j}^{t}+X_{j}^{t})
-1,
\label{nkpz5}
\end{equation}
and 
\begin{equation}
H_{j}^{t+1}
=
\min (H_{j-1}^{t},H_{j}^{t}, H_{j+1}^{t},H_{j}^{t}
+\tilde{X}_{j}^{t})
+1,
\label{nkpz6}
\end{equation}
\eph
In the model (\ref{nkpz5}) a particle is evaporated 
 at the local maxima with the rate $q$ and in the model 
(\ref{nkpz6})  a particle deposited  at the local minima  
with the rate $1-q$.

In the RSOS model  introduced by Kim  and Kosterlitz (KK),\cite{kk}
a particle is deposited at randomly selected site 
as long as the height difference $\Delta h$
between nearest-neighbor columns remains 
as $h\leq 0$.
Then a particle  is deposited in the local minima.
The difference between our B model and the KK model
is time steps.
In the B model 
in one time step 
$m$ particles  are deposited.
It corresponds to $m$ time steps in KK model.
The difference does not affect the universality.


\section{Concluding Remarks}

We consider the ultra-discrete Burgers equation.
We classify the equation into five regions 
in the parameter space.
We discussed the behavior of the solutions.
The A and B models  have  the rectangular solutions.
On the other hand the C model has only shock wave solutions.
We construct the deterministic surface growth  models using 
A and B models.
Furthermore we introduce the noise into the 
 ultra discrete Burgers  equation.
We present the automata models  of KPZ equation.
The A model  becomes the discrete time version of ASEP and 
B model corresponds to the KK  model.
It is well known  ASEP and KK model 
belong to the universality of KPZ equation.
We discussed the symmetry of the ultra-discrete 
and continuous Burgers equation.
We  showed  why automata model represent the 
property of continuous Burgers equation.
Furthermore we study the stability of equations.
Using this result we
can understand the
parts of the meaning of ultra-discrete limit.
  
We can  extend these method to the higher dimension and 
we can introduce the quenched noise into the
ultra-discrete KPZ equation.
About these  contents 
we will  discuss  else where.




\begin{thebibliography}{99}

\bibitem{b}
J. M. Burgers, 
{\it The nonlinear Diffusion Equation}
 (Riedal, Boston , 1974).

\bibitem{kpz}
M.Karder, G.Parisi, and  Y. C. Zhan,
Phys. Rev.  Lett. {\bf 56} 889 (1986). 

\bibitem{kk}
J. M. Kim and J. M. Kosterlitz, 
Phys. Rev. Lett. {\bf 62}  2289 (1989).

\bibitem{n}
T. Nagatani,
Phys. Rev. E {\bf 58} 700 (1998).

\bibitem{nt}
N. Nishinari and D. Takahashi,
J. Phys. A {\bf 31} 5439 (1989).




\bibitem {tt}
T. Tokihiro, D.Takahashi, J. Matsukidaira, and J.Satsuma,
 Phys. Rev. Lett. {\bf 76} 3247 (1996).



\bibitem{ks}
J. Krug and  H.Spohn, 
Phys. Rev. A  {\bf 38} 4271 (1988).

\bibitem{w}
S. Wolfram, 
Rev. Mod. Phys. {\bf 55}  601 (1983).

\bibitem{l}
T.Liggett, 
{\it Interacting Particle Systems}
 (Springer-Verlag, Berlin 1985).
 













\end{thebibliography}
\end{document}